\documentstyle[12pt, fleqn]{article}
\def\ref{\par\noindent\hangindent=6mm\hangafter=1}

\begin{document}
\baselineskip=8mm

\begin{center}

{\bf Typical Scales in the Spatial Distribution of QSOs}

\vspace{35mm}

{\bf Zugan Deng}$^{2}$, \hspace{5mm} {\bf Xiaoyang Xia}$^{3}$,
\hspace{5mm} {\bf Li-Zhi Fang}$^{1}$

\end{center}

\bigskip

\baselineskip=5mm

\begin{description}

\item[\ $^{1}$] Physics Department and Steward Observatory, University of
Arizona, Tucson, Arizona 85721

\item[\ $^{2}$] Graduate School, Chinese Academy of Sciences, Beijing,
  P.R.China

\item[\ $^{3}$] Physics Department, Tianjing Normal College, Tianjing,
     P.R.China

\end{description}

\vspace{100mm}

\begin{center}

To appear in the Astrophysical Journal  August 10, 1994

\end{center}

\newpage
\baselineskip=5mm

\begin{center}

{\bf Abstract}

\end{center}

We present results of searching for the possible typical scales in the
spatial distribution of QSOs. Our method is based on the second
derivative of the two-point correlation function. This statistic
is sensitive to the scale of the maximum in the spectrum $P(k)$ of the
density perturbation in the universe. This maximum or bend scale can be
detected as the wavelengths of the periodic component in the second
derivative of the integral correlation function. For various QSO samples
compiled from surveys of pencil-beam and bright QSOs, a typical scale of
about 93 $\pm$ 10 h$^{-1}$Mpc for $q_0=0.5$ has been detected.
This typical scale is in good agreement with that found in the spatial
distributions of  galaxies, clusters of galaxies, and CIV absorption
systems of QSOs if $q_0$ is taken to be $\sim 0.2$. Therefore, it is likely
a common or universal scale in the large scale structure traced by these
objects. This result is consistent with the assumption that the typical
scale comes from a characteristic scale in the spectrum of the density
perturbation in the universe.

\vspace{5cm}

{\it Subjects headings;} cosmology - QSO: clustering


\newpage

\noindent{\bf 1. Introduction}

\bigskip

According to the standard scenario of structure formation in the
universe, the initial perturbation produced by quantum
fluctuation of scalar fields during the inflationary era is
scale-invariant.
The power spectrum of the initial perturbation is assumed to be
$P(k) \propto k^n$, where $k$ is the wavenumber of the perturbation,
and the spectral index $n \sim 1$. Therefore, no typical scales exist
in the very early universe. Subsequent evolution of the universe
leads to a deviation of the density perturbation spectrum from a
scale-invariant one. Typical scales emerge from the distribution of cosmic
matter. For instance, in a linear regime, the density spectrum can be
approximated as the following form (Peacock 1991; Mo et al 1993)
\begin{equation}
P(k)=\frac{k}{1+(k\lambda/2\pi)^{2.4}}
\end{equation}
where $\lambda$ is a typical scales, on which  $P(k)$ is the maximum, i.e.
$P(k)$ bends from $\propto k$ to $ k^{-1.4}$ at $k\sim 2\pi/\lambda$.
The clustering of galaxies
and clusters of galaxies showed that the bend scale $\lambda$ should be
larger than about 100 h$^{-1}$Mpc. On the other hand, the anisotropy of
cosmic background radiation indicates that $\lambda <$ 1000 h$^{-1}$Mpc.
Therefore, in the standard model, the linear evolution of density
perturbation brings out at least one typical scale in the range between
100 and 1000 h$^{-1}$Mpc in the spectrum.

The possible existence of typical scales in non-standard scenario of
the structure formation has also been proposed. For a modified
inflation model (Starobinsky, 1992), it has been found that a typical
scale of about 100 h$^{-1}$Mpc in the cluster-cluster correlation is
crucial in determining the peculiarities of the inflation and the nature
of the dark matter (Kotok et al 1993).

Observations have indeed discovered structures in the distribution
of galaxies or clusters with scales as large as about 100 h$^{-1}$Mpc,
including the great void (Kirshner et al. 1981), filaments and sheets
(Haynes and Giovannelli, 1986), the Great Wall (de Lapparent et al. 1988),
the Great Attractor (Dressler et al. 1987) and the 128 Mpc `periodicity'
of pencil beam sample (Broadhurst et al. 1990). However, these observed
scales cannot be identified as the bend scale in the density spectrum.

In the last two years, systematic approaches to the typical scales in the
spatial distribution of galaxies and clusters have been done by several
groups. Buryak, et al. (1991, 1992) developed a method to probe typical
scale from one-dimensional samples. Einasto and Gramann (1993) investigated
the possible observational phenomena in the distribution of clusters and
galaxies related to the bend scale in $P(k)$. An extensive search for
the typical scales has been made by Mo, et al. (1992a, b). Using the method
of the second derivative of the integral two point correlation function, they
have detected typical scales in the distribution of galaxies and clusters,
especially, a scale of 130 h$^{-1}$Mpc commonly exists in samples including
the deep pencil-beam survey (Broadhurst, et al. 1990), deep redshift surveys
of Abell clusters (Huchra, et al. 1990) and QDOT survey of IRAS galaxies
(Rowan-Robinson, et al. 1990).

In this paper we extend this investigation to QSOs, i.e. searching for
the possible typical scales in the spatial distribution of QSOs. Our
motivations are twofold. First, the distributions of low-redshift objects
like galaxies showed that the bend in the density spectrum may occur
at wavelengths $\lambda \sim$ 150 h$^{-1}$Mpc (Peacock 1991; Vogeley, et al.
1992; Peacock \& West 1992; Mo, et al. 1992a,b;
Jing \& Valdarnini 1993). However, the wavelengths involved are already
comparable to the sizes of the samples used, and the fair-sample assumption
may then be questionable. This problem should be less severe for QSOs
because we can have QSO samples with size much greater than 150 h$^{-1}$Mpc.
Secondly, if the typical scales detected in the structures of galaxy and
cluster do come from the characteristic scales like the bend in the
spectrum of perturbation, it should be measurable in the QSO distribution
as well, unless QSOs trace substantially different large-scale structures
than galaxies and clusters do. Therefore, it is important to see if the
distribution of QSOs is consistent with the assumption that the typical
scales found in galaxies and clusters are `universal'.

It is the interesting to note that the scale 100 h$^{-1}$Mpc has already been
mentioned in early studies of QSO clustering. About a decade ago, using
the nearest neighbor analysis, Chu and Zhu (1983) showed that the
distribution of QSOs listed in the sample Bolton and Savege (1979) deviates
from the Monte Carlo samples on the scale of about 100 h$^{-1}$Mpc. Some
authors also suggested the existence of isolated groups with comoving scales
of about 100 h$^{-1}$Mpc (e.g., Crampton, et al. 1989; Clowes \& Campusano
1991). But these results do not provide a convincing argument for the scale
considered.
We will use more rigorous statistic to detect the typical scales in the
samples of Boyle, et al. (1990, 1991) and Foltz, et al. (1987, 1989).

Our plan is to give a brief description of the method in section 2,
the results of typical scale analysis of QSO samples in section 3,
a comparison of the typical scales of QSOs with that of galaxies in
section 4, and a conclusion in section 5.

\bigskip

\noindent{\bf 2. Method of Detecting Typical Scales}

\bigskip

Statistics based on the amplitude of the two point correlation function
$\xi(r)$ is the most popular method in the study of large scale structure.
This method is, however, not adequate for detecting typical scales.
The amplitude and the correlation length $r_0$ in the two-point correlation
function $\xi(r) = (r/r_0)^{-1.8}$ do not relate to the bend scale
in a simple way. For a given density spectrum $P(k)$, the two point
correlation function can be calculated by
\begin{equation}
\xi (r) = \frac{B}{r}\int_0^\infty \sin kr P(k) kdk
\end{equation}
where B is a constant. Generally, the bend in the spectrum $P(k)$ only
leads to a slight drop in the amplitude of the correlation function on
the scale of bending. This means, only the amplitudes of $\xi(r)$ at
$r \sim \lambda$ are useful to probe $\lambda$. However, on such large
scales, the absolute value of the amplitudes of the two-point correlation
function has a large statistical error due to the uncertainty in
the mean density of objects considered. Structures on larger scales
with density contrast less than
the uncertainty of the mean density will be masked by the noise of the
two-point correlation function. This problem is especially severe for QSO
samples because the mean density of QSOs is redshift-dependent. Even with
a homogeneous sample, it is still difficult to calculate the evolution of
the mean volume density, because the deceleration parameter $q_{o}$ is
poorly determined.

The method developed by Mo, et al. (1992a,b) and Einasto and Gramann (1993)
is based on the second derivative of the two-point correlation function.
We will introduce this method by a slightly different way in order to
demonstrate its advantage in searching for the typical scale, especially
the bend scale in $P(k)$. Let us consider
the behavior of $\xi(r)$ when $r$ is large. Eq.(2) shows that, for a
spectrum with a maximum like that in eq.(1), the dominate term of $\xi(r)$
when $r \geq
\lambda$ should be a periodic function of $r$ with wavelength equal to
about $\lambda$. For instance, if one takes an approximate form of
the spectrum (1) as follows: $P(k)= k$ for $k< 2\pi/\lambda$ and $P(k) =
k^{-1.4}$ for $k> 2\pi/\lambda$, the dominant term of $\xi(r)$ at large $r$
will be
$r^{-m}\cos(2\pi/\lambda r)$, where $m \sim 2$. The second derivative of
$\log\xi(r)$ is then proportional to $\cos(2\pi r/\lambda)$. Therefore,
the bend scale $\lambda$ can be detected by the wavelengths of the
periodic components in the second derivative $d^2\log\xi(r)/dr^2$. Of
course, such periodic components will also be masked by the noise given by
the uncertainty of the mean density. However, it is well known from
statistics that, for a given noise masked data set, identifying periodic
components is easier than determining the absolute value of the
amplitudes of the correlation function. Considering
the sizes of QSO samples usually are much greater than the wavelengths of
the periodic components involved, the statistic of detecting periodic
component in a QSO sample would be more effective than that of
determining the amplitude.

In actual work, the usual two-point correlation function $\xi(r)$
is replaced by the function $\Xi(r)= 1+ \bar\xi(r)$, where $\bar\xi(r)$
is the integral two-point correlation function defined by
\begin{equation}
\bar{\xi} (r)=\frac{3}{r^3} \int _0^r \xi (x)x^2 dx
\end{equation}
When $r \geq \lambda$, $\Xi(r)$ has about the same behavior as $\xi(r)$.
Therefore, we can also use the statistic of $d^2\log\Xi(r)/dr^2$ to
detect the typical scales. In measuring clustering of high redshift objects
on large scales, the statistic based on the integral two point correlation
function $\bar{\xi}(r)$ is sometimes more advantageous than $\xi(r)$.
The reasons are as follows.

First, for determining the two-point correlation function $\xi(r)$, one
needs a choice of bin size of the separation of QSO pair. The binning
may lead to false periodic components in the $\xi(r)$ with wavelengths
equal to the harmonics of the bin scale, and then a misidentification
of the typical scales. Moreover, the total number and number density in
available samples of QSOs are very low, and the binning will cause a large
fluctuation in $\xi(r)$ if the bin scale is chosen too small. These
puzzles can be avoided by using the statistic $\bar{\xi}(r)$ because,
according to the definition eq.(3), it does not bin.

Second, from eqs. (2) and (3), $\bar{\xi}(r)$ can be related to the
density spectrum by
\begin{equation}
\bar{\xi}(r)=\frac{3B}{r^3}\int_0^\infty [\sin kr -kr\cos kr]
P(k) \frac{dk}{k}
\end{equation}
where the window function $[\sin kr- kr\cos kr]/k$ dies off faster at
large $k$ than that for $\xi(r)$ [eq.(2)]. Therefore, the statistical
result will be less severely affected by clustering on small scales
(Mo, et al. 1993).

When the boundary effect is negligible,  $\Xi(r)$ is given by
\begin{equation}
\Xi(r)= \frac{N_{dd}(r) \times N_r}{N_{dr} \times N}
\end{equation}
where $N_{dd}(r)$ is the number of QSO pairs with separation less than
$r$, $N_{dr}(r)$ is the mean number of object pairs between observed
and random samples, $N$ and $N_{r}$ are the total numbers of objects
in real and random samples, respectively.

Eq.(5) shows that $\Xi(r)$ is given by an un-normalized integrated
pair counts of QSOs, the result does not sensitively depend on the
mean number density of QSOs. Therefore, the uncertainty in the mean
density is avoided from the beginning. As a consequence, this method
is not sensitive to the selection function used for generating the random
samples as well. In this paper, we will fit one-dimensional samples
by a cubic polynomial, and three-dimensional samples by a linear
function in the redshift range of each sample.

In calculating the derivative of $\Xi(r)$, we will meet differences like
$N_{dd}(r+\Delta r) - N_{dd}(r)$. Obviously, this difference will be
dominated
by noise when $\Delta r$ is less than the mean distance $D$ of nearest
neighbor
of QSOs in the sample. This will be the main source of the error in the
derivative of the correlation function when $\Delta r \leq D$.
In order to
suppress the influence of this noise in small wavelengths, we smooth
$\Xi(r)$ by convolution
integral $\bar{\Xi}(r)=\int \Xi(r')S(r-r')dr'$, where the smoothing
function $S(r)$ is equal to 1 when
$|r-r'|<L$, and 0 otherwise, and taking the smooth scale $L$ to be equal
to or less than $D$.
Fluctuations with wavelengths less than the scale $L$ will totally be
suppressed in the function $\bar\Xi(r)$ by the smoothing, while all
inhomogeneities with scales comparable to or larger than the scale $L$ will
not be affected by the smoothing. Our algorithm is to use this smoothed
function $\bar\Xi(r)$ to calculate the second derivative
$\Delta\theta(r) \equiv d^2\log\bar\Xi(r)/dr^2$.

The statistical significance of the peaks in the second derivative
$\Delta\theta(r)$
can be measured by the standard deviation $\sigma$ which is
estimated by Monte Carlo samples generated under the same selection
conditions as the real samples. Usually we take 100 random samples to
calculate the standard deviation. Comparing the curve $\Delta\theta(r)$ of
the real samples with that of random samples, we can infer the statistical
significance of peaks appearing in the $\Delta\theta(r)$ of real samples.
Estimating the significance in this way has the advantage that the edge
effects are automatically avoided.

The periodic components in $\Delta\theta(r)$ can be detected by power
spectrum analysis (PSA). The wavelengths of these periodic components are
the typical scales.  The statistical significance of the existence of
periodic components in the second derivative, $\Delta\theta(r)$,
can be estimated by the usual way of power spectrum analysis.

This method has been used to analyze 1- and 3-dimension samples of optical
and IRAS galaxies or clusters of galaxies, and a common scale of
130 $\pm 10$ h$^{-1}$Mpc
was detected (Mo, et al. 1992a,b). The samples used for analysis include
the deep pencil-beam
surveys (Broadhurst et al. 1990), deep redshift surveys of Abell clusters
(Huchra et al. 1990) and QDOT survey of IRAS galaxies (Rowan-Robinson et al.
1990). Therefore, the scale of 130 h$^{-1}$Mpc might be a candidate for the
bend scale $\lambda$ in the initial density spectrum $P(k)$. Since the
structures with scales as large as about 100 h$^{-1}$ Mpc in the present
universe should still remain in the linear evolutionary stage, the typical
scale found in the distribution of galaxies and clusters should probably also
be measurable in the distributions of high redshift objects. Therefore,
one should expect the existence of a 100 h$^{-1}$Mpc typical scales in QSO
distribution if QSOs trace the high peaks in the density field as galaxies
and clusters of galaxies do.

\bigskip

\noindent{\bf 3. Statistical Results}

\smallskip

\noindent {\bf 3.1 Pencil-beam samples}

\smallskip

The one-dimensional samples of QSOs used in our analysis are formed
from Boyle et al 1990 (BFSP) and 1991 (BJS). BFSP
contains about 420 QSOs identified in a complete (to $B \leq 21$),
ultraviolet excess (UVX) survey, which covers 34 pencil beam fields,
each has about 0.35 square degrees. These pencil-beam fields are
scattered over eight $5^{\circ}\times5^{\circ}$ UK Schmidt fields.
BJS includes  61 QSOs identified in a complete to $B \leq 22$ survey
done by multicolored technique in three pencil-beam fields at high
galactic latitudes.

Redshift distribution of QSOs listed in BSFP and BJS
are plotted in Figure 1a and b, respectively. As well known, the UVX and
multicolored technique are likely to provide QSO candidates with high
completeness when redshift $z \leq 2.2$ (V\'eron 1983). On the other hand,
the imposed stellar morphological criterion may cause the incompleteness
at redshift less than about 0.6. Figure 1 shows that most QSOs in the BFSP
and BJS are in the redshift range of 0.6 to 2.2 and the number of QSOs
dramatically decreases outside this interval. Therefore, the samples
consisting of QSOs with redshifts from 0.6 to 2.2 in BSFP and BJS
should be largely complete and unbiased. We adopt, respectively, QSOs
with $0.6 < z <2.2$ in 1)  BSFP and 2) BJS as two parent samples in our
statistic.

Because each field covers only about 0.35 $deg^{2}$, the size of their
cross section is about 20 h$^{-1}$Mpc at $z\sim 1$. One can consider these
samples as one-dimensional if we focus on the structures with scales much
larger than 20 h$^{-1}$Mpc. Each one-dimensional sample can be seen as a
representation of the three-dimensional distribution in a given direction.
Obviously, some features shown in these samples are
direction-dependent. In order to reduce the influence of the local features
on the statistics, we use combined sample, which consists of a number of
the pencil-beam samples in different directions. For such combined samples,
the individual features of the pencil-beam fields should be less important.

We made four subsamples called QN, QS, QSGP and QBJS, which consist of
QSOs in
the fields of northern sky, southern sky (excepted those in the Schmidt
field along southern galactic pole direction), southern galactic pole
and that given by BJS survey, respectively. Table 1 shows the numbers
of pencil-beam fields $N_p$, numbers of QSOs $N_q$ for each subsample.
It should be pointed out that a Schmidt field is about $5^{\circ} \times
5^{\circ}$, two pencil-beam fields in the same Schmidt plate have a mean
separation of about 2$^{\circ}$.5 which corresponds to an across scale of
about 90 h$^{-1}$Mpc at $z\sim 1$. Thus, in studying the structures with
scales larger than 100 h$^{-1}$Mpc, the pencil-beam samples in the same
Schmidt plate should not be considered as totally independent samples,
i.e. different pencil-beams may imprinted by a same structure on scale
larger than the separation of the pencil-beams.

Figures 2a-d present the results of the second derivative,
$\Delta\theta(r)$, for each
subsample. The bold lines show the $\Delta\theta(r)$ for the real samples,
and the light lines are the standard deviations $\pm 1\sigma$ given by random
samples. Since the mean distance of nearest neighbor QSOs in these samples
is equal to or larger than about 50 h$^{-1}$ Mpc, the smoothing scale $L$ is
taken to be 40 h$^{-1}$Mpc. The sharp dips appearing in Figure 2 on scales
less than about 40 h$^{-1}$Mpc are caused by that, for all random samples,
$N_{dr}(r)$ is zero on small scales. It can be clearly seen from Figure 2
that, for all subsamples of QSOs, the curves of $\Delta\theta(r)$ show
periodically distributed peaks (and valleys). Figure 3a and b show the
same calculation as Figure 2 but the smoothing scale $L$ is taken to be
20 h$^{-1}$Mpc. As expected, more fluctuations with short wavelengths
appeared in Figure 3, while the fluctuations on scales larger than
40 h$^{-1}$Mpc are the same as the case of $L =40$ h$^{-1}$Mpc.

The statistical significances of each peak (and valley) in $\Delta\theta(r)$
are marginally higher than $\pm\sigma$, mostly in the range of $1 - 2.5 \
\sigma$. This result is the same as that given by previous studies. Up to
now, almost all QSO structures detected with scales greater than 10
h$^{-1}$Mpc are in the significance level of 2 - 3 $\sigma$ (Ivovino
\& Shaver, 1988; Bahcall \& Chokshi 1991; Boyle \& Mo 1993). However, our
method is not only based on the individual peak in $\Delta\theta(r)$, but
in the regular or periodic distribution of these peaks (and valleys).
Figure 4 and 5 plot, respectively, the power spectrum of the
$\Delta\theta(r)$ for sample QS and QBJS with $L=$ 40 and 20 h$^{-1}$Mpc.
One can find from these figures that 1) all power spectrum show a peak around
$\lambda \sim$ 90 h$^{-1}$Mpc with confidence no less than 99\% (for $L$=
40 h$^{-1}$Mpc); 2) if we use the width of the peak as a measure of the
uncertainty of the wavelength, the mean wavelength for one-dimensional
samples is $90 \pm 8$ h$^{-1}$Mpc; 3) for each subsample the spectrum of
$L=$ 20 and 40 h$^{-1}$Mpc have the same shape, therefore these statistical
results are independent of the smoothing length.

\bigskip

\noindent{\bf 3.2 Three dimensional samples}

\smallskip

The three-dimensional sample used in this paper is compiled from the
LBQS survey (Foltz et al., 1987, 1989; Hewett et al., 1991; Chaffee et
al., 1991; Morris et al., 1991). The LBQS survey presented more than
1000 QSOs with $m_{j} \leq $18.5. The total area of these fields is about
800 square degrees. It is one of the largest and uniformly selected
QSO sample up to date. The redshift interval of this sample is between
0.2 and 3.3. An artificial cut-off has been made at redshift 0.2, because
too many stars mixed into the candidates below this redshift. As the
redshifts of QSOs becomes higher than 3.3, the Ly$\alpha$ line will
move out of the $j$ band. The sample can be considered to be complete
in the magnitude interval $16.0 < m_{j} < 18.7$. Figure 6a plots the
redshift distributions of QSOs in samples LBQS.

The first sample compiled from this survey is called LBQS, which contains
of all LBQS QSOs with redshift in the range from 1.0 to 2.2. The
limitation of $1.0 < z < 2.2$ comes from the following consideration.
a) Each plate used in LBQS survey coves an area of about
$6^{\circ}\times 6^{\circ}$, which  spans $\sim$ 200 h$^{-1}$Mpc and
higher when $z>1$ and $q_0=0.5$. Therefore, if we are
interesting in probing structures with scales equal to or larger than
100 h$^{-1}$Mpc, only the sub-samples with $z>1$ can be treated as a three
dimension one. b) The number of the LBQS QSOs drops rapidly when $z>2.2$.
The mean distance of nearest neighbor QSOs at $z > 2.2$ is equal to or
larger than $\sim$ 100 h$^{-1}$Mpc. Therefore, the data at $z>2.2$ are no
longer
suitable for probing structures with scales of about 100 h$^{-1}$Mpc.

To study the possible influence of the foreground objects on the typical
scales, such as that given by gravitational lensing effect, we
compiled a sample called LBQS-V, which consist of all LBQS QSOs with
$1.0<z<2.2$ excepting those in field of the nearest supercluster Virgo
(Figure 6b). The redshift distributions of both LBQS and LBQS-V are
quite smooth. Samples LBQS and LBQS-V are listed in Table 2.

Figures 7a and b plotted the result of $\Delta\theta(r)$ for the LBQS and
LBQS-V, respectively. The ranges of $\pm1 \sigma$ given by random samples
are shown by the light curves. The ratio of signal
to noise in 3-dimensional samples appears to be higher than that of the
pencil beam samples. The significance of the peaks (and valleys) now is
about 3$\sigma$. As in the 1-dimensional sample, the
significant peaks distributed regularly or periodically. Figures 8 and 9
show, respectively, the power spectrum of samples LBQS and LBQS-V with
$L=$ 40 and 20 h$^{-1}$Mpc. The mean wavelength is $\sim 95 \pm 9$
h$^{-1}$Mpc. This is the same as that of pencil beam samples.

The value of $\lambda \sim$ 90-100 h$^{-1}$Mpc found here is in
agreement with those obtained from the amplitudes of the correlation
function
(Mo \& Fang 1993). Using a best fit of the integral correlation function
to the power spectrum $P(k)$ [eq.(1)], it found $\lambda \sim 100 -200
\ $ h$^{-1}$Mpc. Therefore, the scale of $\sim$ 100 h$^{-1}$Mpc seems to
be universal for the various QSO samples considered.

\bigskip

\noindent{\bf 4. Difference of Typical Scales between QSOs and Galaxies}

\smallskip

The typical scale found in QSO distribution  shows a difference
from that of galaxies and clusters of galaxies (Mo, et al. 1992a),
which was found to be 130 $\pm 10$ h$^{-1}$Mpc by the same method.
It is generally believed that the structures with scales
larger than about 50 h$^{-1}$Mpc should still remain in linear evolution
regime. If the typical scale comes from the bend scale in the initial
density spectrum, the comoving typical scale of QSOs
should be the same as that of galaxies and clusters of galaxies. Therefore,
it is necessary to study the possible origin of the observed
difference
between the typical scales of QSOs, and galaxies and clusters.

If the typical scale is assumed to be local, i.e. being only a feature
of nearby galaxies, one should not expect that the same typical scale
shows up in the spatial distribution of QSOs. However, it has been found
that the typical scale of 130 h$^{-1}$Mpc exist in almost all samples
of galaxies and clusters of galaxies (Mo, et al, 1992a, b).
On the other hand,
the fact that no difference has been found between the results of LBQS and
LBQS-V (Figure 6a and b) indicates that the 95 h$^{-1}$Mpc typical scale
of QSOs may not be affected (at least in current error bar) by gravitation
lensing of a local structures like the Virgo clusters. Therefore, one cannot
explain the difference of the typical scale between galaxies and QSOs as
a local effect.

If galaxies and QSOs trace different aspects of the density fields in
the universe, we should not expect that galaxies and QSOs have the same
typical scales. Considering the bias mechanism for galaxies is probably
no longer useful for QSO formation, one may reasonably assume that the
QSO-traced structures are different
from that traced by galaxy. Indeed, the number density of QSOs is much
less than galaxies. Therefore, in terms of bias model, the biasing
threshold of QSO
should be higher than galaxies. Bower, et al. (1993) recently proposed a
bias model describing cooperative formation of galaxies,
in which the threshold of galaxy formation is scale-dependent, it is
lower in a domain with higher mean density, and higher in the area with
lower mean density. This is equal to replacing $P(k)$ by $P(k)B(k)$, where
the bias function $B(k)$ is decreasing with $k$ increasing. Obviously,
the bend scale of ``spectrum'' $P(k)B(k)$ will be greater than that of
$P(k)$.

However, this explanation encounters difficult if we consider
the following facts. First, QSO clustering satisfies the same power law
correlation
function as galaxies and clusters (Shanks et al. 1988; Fang et al. 1985;
 Shaver 1988;
Chu \& Zhu 1989; Crampton, Cowley \& Hartwick 1989; Boyle 1991).
Second, QSO clustering is the same as small groups of galaxies or poor
clusters (Bahcall \& Chokshi 1991). It has been known for a decade
that low redshift QSOs are
preferentially located in poor clusters or groups of galaxies. This was
found by the QSO-galaxy covariance function (Yee \& Green 1987),
CIV-associated absorption in high redshift radio-loud QSOs (Flotz et al
1988), clustering analyses of the QSO distribution and galaxy environments
around of QSOs (Ellingson et al 1991a). It has also been shown that the
velocity dispersion of galaxies around QSOs
is $\rm \sim 400 \, km \, {\rm s}^{-1}$ (Ellingson et al 1991b). This means,
QSOs trace the same density field as poor clusters do. Therefore, the
formation and radiation of QSOs may not provide an effective mechanism that
leads to the difference of typical scales between QSOs and clusters of
galaxies.

Now we turn to the explanation based on $q_0$-dependence of the typical
scale. In the previous sections, all scales are calculated under the
assumption that the universe is of Einstein-de Sitter, and thus the
deceleration parameter $q_0$ is taken to be 0.5. As it has been pointed out
by Shank et al (1987) and Mo, et al. (1992a, b), the typical scale is
crucially dependent on $q_0$ for high redshift objects like QSOs.
Figures 10 is the power spectrum of the $\Delta\theta(r)$ for QBJS
when $q_0$ $=$ 0.2, 0.4 and 0.7, respectively. From the peaks in Figure
10 one can see a systematic increase of the wavelength with the decrease of
$q_0$. This relationship is plotted in Figure 11. It shows that when
$q_0=0.2$, the typical scale of QSOs is 125 $\pm$ 11 h$^{-1}$Mpc. For
other samples, we found about the same result. Figure 12 and 13 showed
that all power spectrum of samples QN, QSGP, LBQS and LBQS-V have a peak
at 130 $\pm$ 15 h$^{-1}$Mpc.

Since galaxies and clusters have low redshifts, their typical scales do
not depend on $q_0$. Therefore, the typical scale of QSOs is in good
agreement with that of galaxies and clusters (Mo, et al. 1992a, b) when
$q_0 \sim 0.2$. In other words, if one assumes that the comoving value
of the typical scale can be used as a ``standard cosmological rod'' in
the linear regime of an expanding universe, the universe should be of
$q_0 \sim 0.5$. Considering various uncertainties in the typical scales
of galaxies, clusters, and QSOs, it would be better to say that $q_0$
should be less than 0.5 at $95\%$ confidence.

It is interested to point out that the conclusion of a low total mass density
universe ($q_0 \sim 0.1$) has also been proposed by several independent
researchers (Park, et al. 1992; Bahcall and Cen, 1992; Vogeley, et al., 1992).
In the paper by Mo, et al (1992b), the $q_0$-dependence of the typical scales
of the CIV absorption systems and the Ly$\alpha$ forests of high redshift
QSOs has been studied. Using their result, one can find that the value
of $q_0$, at which the typical scale of CIV system is the same as that of
galaxies and cluster, is $q_{0}\sim 0.2-0.3$. Therefore, when $q_0=0.2$,
the 130 h$^{-1}$Mpc typical scale is likely universal among the samples of
 galaxies,
clusters, QSOs and the CIV  absorption systems of QSOs. However, for this
$q_0$, the
typical scale of Ly$\alpha$ forest is different from 130 h$^{-1}$Mpc. This
is understandable because the CIV systems probably originated
from absorption of clouds associated with galaxies (Waymann, et al. 1979;
Young et al, 1982), and Ly$\alpha$ comes from clouds which are unable to
form QSOs and galaxies. In a word, it would be reasonable to say that the
130 h$^{-1}$Mpc typical scale is universal in the distributions of galaxies,
clusters and quasars.

\bigskip

\noindent{\bf 5. Conclusion and Discussion}

\smallskip

The possible typical scales in the distribution of QSOs have been
detected by means of the second derivative of integrated correlation
function.  A typical scale of about 93 $\pm$ 10 h$^{-1}$Mpc when $q_0=0.5$
have been detected with considerable confidence in available 1- and
3-dimension samples of QSOs. This typical scale is probably ``universal''
for various subsamples of QSOs. If $q_0$ is taken to be 0.2, the QSO
typical scale becomes the same as that of galaxies and clusters
of galaxies. One can then have the following conclusions.

1. The existence of a $\sim$ 100 h$^{-1}$Mpc typical scale in the spatial
distribution of QSOs further strengthens the picture that QSOs probably
trace the same larger scale density field of the universe as galaxies and
clusters of galaxies do.

2. The redshifts range and sizes of QSO samples used here are totally
different from that of galaxies and clusters of galaxies. The agreement of
the QSO typical scale with the result of galaxies and clusters suggests
that the detected typical scale is most likely universal in the large scale
structure. This is consistent with the assumption that the typical scale
is due to a characteristic scale in the initial perturbation spectrum,
such as the bend scale in the density spectrum at the linear regime.

3. If the typical scale comes from the initial spectrum of the density
perturbation, one can use the comoving values of the typical scale as a
cosmic standard length when this scale remains in linear evolution.
Accordingly, if we require that the comoving value of typical scale of
QSOs is the same as that of galaxies, clusters and CIV absorption systems,
the typical scale measurement favors an open universe, i.e. $q_0 < 0.5$.

4. The bend scale is model-dependent. For instance, CDM model should
have a lower bending scale than that of a hybrid model. The formation time
of structures with scale as large as the bend scale is also model-dependent.
Therefore, the fact that the structures of bend scale exist at the high
redshift seen for QSOs distribution may help discriminate among various
cosmological models.

\bigskip

The authors thank Drs. Hou-jun Mo, and Yipeng Jing and Mr. Jesus Pando
for useful discussion.
Deng is grateful to the University of Arizona for hospitality and support
from the international exchange program. D and X thank also the support
of Chinese NSF. This work was also partially supported by NSF contract
INT 9301805.

\newpage

\begin{center}

	Table 1

 Data of  1-dimensional samples

\vspace{5mm}
\begin{tabular}{llrl}
\hline
Sample    &  $N_p$    &  $N_q$  & Notes\\
\hline
          &          &        &          \\
   QN     &      13  &  141   &  northern sky  (BSFP)\\

   QS     &      14  &  154   & southern sky excepting\\
          &          &        & southern galactic pole (BSFP)\\
  QSGP    &       7   &  94   &  southern galactic pole (BSFP)\\

  QBJS    &       3   &  52   &   (BJS)\\
          &           &       &      \\
\hline
\end{tabular}

\end{center}

\vspace{10mm}

\begin{center}

 Table 2

 Data of 3-dimensional samples

\vspace{5mm}

\begin{tabular}{llrl}
\hline
Sample   & $N_p$  & $N_q$ &   Notes \\
\hline
         &        &       &         \\

LBQS     &   18   & 510   &     Foltz et al, 1987, 1989; Hewett et al. 1991
and\\
         &        &       &      Chaffee et al., 1991, Morris et al. 1991\\
LBQS-V   &   14   & 399   &      excepting Virgo fields \\
         &        &       &      (references are the same as LBQS)\\
         &        &       & \\
\hline
\end{tabular}

\end{center}

\newpage

\noindent{\bf References}

\bigskip

\ref Bahcall N.A. \& Cen R. 1992, ApJ, 398, 281

\ref Bahcall N.A. \& Chokshi A. 1992, ApJ, 380, L9

\ref Bolton J.G. \& Savage A. 1979, MNRAS, 188, 599

\ref Bower R.G., Coles P., Frenk C.S. \& White S. 1993, ApJ, 405, 403

\ref Boyle B.J., Fong R., Shanks T. \& Peterson B.A. 1990,
MNRAS, 243 ,1  (BFSP)

\ref Boyle B.J., Jones L.R. \& Shanks T. 1991, MNRAS, 251, 482 (BJS)

\ref Boyle B.J. 1991, Proc. of Texas/ESO -CERN Symp. on Relativistic
Astrophysics, Cosmology and Fundamental Physics, 14

\ref Boyle B.J. \& Mo H.J. 1993, MNRAS, 260, 925

\ref Broadhurst T., Ellis R., Koo D. \& Szalay A. 1990, Nature, 243, 726

\ref Buryak O.E., Demiansky M. \& Doroshkevich A.G. 1991, ApJ, 383, 41

\ref Buryak O.E., Demiansky M. \& Doroshkevich A.G. 1992, ApJ, 393, 464

\ref Chaffee F.H. et al. 1991: AJ, 102, 461

\ref Chu Y.Q. \& Zhu X.F. 1983, ApJ, 267, 4

\ref Chu Y.Q. \& Zhu X.F. 1988, A\&A, 215, 14

\ref Clowes R.G. \& Campusano L.E. 1991, MNRAS, 249, 218

\ref Crampton D., Cowley A.P. \& Hartaick F.D.A. 1989, ApJ, 345, 59

\ref de Lapparent V., Geller M.J. \& Huchra J.P. 1988, ApJ, 332, 44

\ref Dressler A., Lynden-Bell D., Burstein D., Davis R.L., Faber S.M.,
Terlevich R.J. \& Wegner G. 1987, ApJ, 313, 42

\ref Einasto J. \& Gramann M. 1993, ApJ, 407, 443

\ref Ellingson E., Yee H.K.C. \& Green R.F. 1991a, ApJ, 371, 49

\ref Ellingson E., Green R.F., \& Yee H.K.C. 1991b, ApJ, 378, 476

\ref Fang L.Z., Chu Y.Q. \& Zhu X.F. 1985, Ap\&SS, 115, 99

\ref Foltz C.A. et al., 1987, AJ, 94, 1423

\ref Foltz C.B., Chaffee F.H., Weymann R.J. \& Anderson S.F. 1988,
     in {\it QSO Absorption Lines}, eds. Blads, J.C., Turnshek, D. \& Norman,
     C.A., Cambridge\

\ref Foltz C.A. et al. 1989, AJ, 98, 1959

\ref Haynes M.P. \& Giovanelli R. 1986, ApJ, 306, L55

\ref Hewett P.C. et al. 1991, AJ, 101, 1121

\ref Huchra J., Henry J., Postman M. \& Geller M. 1990, ApJ, 365, 66

\ref Iovino A. \& Shaver P.A. 1988, ApJ, 330, L13

\ref Jing Y.P. \& Valdarnini R. 1993, ApJ, 406, 6

\ref Kirshner R.P., Oemler A., Schechter P. \& Schectmen S.A. 1981,
 ApJ, 248, L57

\ref Kotok E.V., Naselsky P.D. \& Novikov D.I. 1993, preprint of NORDITA

\ref Mo H.J. \& Fang L.Z. 1993, ApJ, 410, 493

\ref Mo H.J., Xia X.Y., Deng Z.G., B\"orner G. \& Fang L.Z. 1992a, AA,
256, L23

\ref Mo H.J., Deng Z.G., Xia X.Y., Schiller P. \& B\"orner G. 1992b,
AA, 257, 1

\ref Mo H.J., Peacock, J.A. \& Xia, X.Y. 1993, MNRAS, 260, 121

\ref Morris S.L. et al. 1991, AJ, 102, 1627

\ref Park C., Gott III J.K. \& Da Costa L.N. 1992, ApJ, 392, L51

\ref Peacock J.A. 1991, MNRAS, 253, 1p.

\ref Peacock J.A. \& West, M. 1992, MNRAS, 259, 494

\ref Rowan-Robinson R. et al. 1990, MNRAS, 247, 1

\ref Shanks T., Fong R., Boyle B.J. \& Peterson B.A. 1987, MNRAS,
     227, 739

\ref Shanks T., Boyle B.J. \& Peterson B. 1988, in Proceedings
of a Workshop on Optical Surveys for Quasars, ed. Osmer et al.
(ASP Conf. Ser., 2), 244

\ref Shaver P. 1988, in IAU Symp. 130, Large-Scale Structure
of the Universe, ed. J.Audouze et al. (Dordrecht: Kluwer), 359

\ref Starobinsky A.A. 1992, JETF Lett, 55, 489

\ref V\'eron P. 1983, in "QSOs and Gravitational Lenses", Proc.
24th Liege

\ref Vogeley M.S. et al. 1992, ApJ, 391, L5

\ref Weymann R.J., Williams R.E., Peterson B.M. \& Turnshek D.A.
1979, APJ, 234, 33

\ref Yee H.K.C. \& Green R.F. 1987, ApJ, 319, 28

\ref Young P., Sargent W.L.W. and Boksenberg A. 1982, ApJS, 48, 455

\newpage

\noindent{\bf Figure captions}:

\bigskip

\begin{description}

\item[Figure 1]  Redshift histograms of QSOs listed in pencil beam
     surveys. a. BFSP given by Boyle et al. (1990), and b. BJS given by
     Boyle et al. (1991).

\item[Figure 2] Curves of $\Delta\theta(r)$ of samples a) QN; b) QBJS;
      c) QS; and d) QSGP. Light lines show the curves of $\pm\sigma$
      given by a average of 100 random samples. $q_0$ is taken to be 0.5,
      and smoothing scale $L =$ 40 h$^{-1}$Mpc.

\item[Figure 3] Curves of $\Delta\theta(r)$ of samples a) QN and b) QS.
      These curves are obtained in the same way as Figure 2, but taking
      the smoothing scale $L =$ 20 h$^{-1}$Mpc.

\item[Figure 4] Power spectrum of $\Delta\theta(r)$ for sample QS. The
       smoothing length $L$ is taken to be 40 and 20 h$^{-1}$Mpc, and
       $q_0=0.5$. For peaks with $P \geq 7.6$, the confidence of the
       existence of a periodic component is $\geq$ 99\%.

\item[Figure 5] Power spectrum of $\Delta\theta(r)$ for sample QBJS.

\item[Figure 6] Redshift distributions of QSOs listed in the LBQS
      survey (Foltz et al. 1988, 1989): a. LBQS, consisting of all LBQS QSOs;
      b. LBQS-V, consisting of
      all LBQS QSOs, but excepting those in the area of the Virgo cluster.

\item[Figure 7] Curves of $\Delta\theta(r)$ of three-dimensional
      samples, a. LBQS; b. LBQS-V, respectively. $q_0$ is taken to be 0.5.

\item[Figure 8] Power spectrum of $\Delta\theta(r)$ for sample LBQS.

\item[Figure 9] Power spectrum of $\Delta\theta(r)$ for sample LBQS-V.

\item[Figure 10] Power spectrum of $\Delta\theta(r)$ for sample QBJS.
      The deceleration parameter is taken to be $q_0=$ 0.2, 0.4, 0.7,
      respectively.

\item[Figure 11] QSO's Typical scale in sample QBJS as a function of
    $q_0$.

\item[Figure 12] Power spectrum of $\Delta\theta(r)$ for samples QN and
   QSGP when $q_0=0.2$ and $L=$ 40 h$^{-1}$Mpc.

\item[Figure 13] Power spectrum of $\Delta\theta(r)$ for samples LBQS
   and LBQS-V when $q_0=0.2$ and $L=$ 40 h$^{-1}$Mpc.

\end{description}

\end{document}